\documentclass[10pt]{amsart}
\usepackage{latexsym}
\usepackage{amsmath}
\usepackage{amssymb}

\def\IF{\operatorname{\fam0 IF}}
\def\MCQ{\operatorname{\fam0 MCQ}}

\begin{document}
\title
{THE GAME OF CIPHER BEADS}
\date{May 20, 2009}
\author{S.~S. KUTATELADZE}
\address[]{
Sobolev Institute of Mathematics\newline
\indent 4 Koptyug Avenue\newline
\indent Novosibirsk, 630090\newline
\indent RUSSIA
}
\begin{abstract}
Comparison  between the various impact factors of
a few Russian journals demonstrates the deficiencies
of the popular citation indices.
\end{abstract}
\email{
sskut@member.ams.org
}
\maketitle

Since recently there has been much ado invoked in science
by incessant attempts at replacing expertise
with numerical manipulations.
Of especial relevance to the Russian mathematical community are
the following indices:

\begin{itemize}

\item[$\bullet$]
MCQ,
the

Mathematical Citation Quotient

of the American Mathematical Society which utilizes the database of

{\it Mathematical Reviews}
 (abbreviated to~MR);

\item[$\bullet$]
IF or ISI,
the classical

impact factor

of the

Institute for Scientific Information

(which is a part of the

Thomson Reuters Corporation);

\item[$\bullet$]
RISC,
the

Russian Index of Scientific Citation\footnote{Cp.~\cite{RISC}}

which rests upon the
database of the

Scientific Electronic Library;

\item[$\bullet$]
{MNRU, the impact factor of the

All-Russia Mathematical Portal

Math-Net.Ru
which uses its own database.}\footnote{Cp.~\cite{MNRU}.}
                                      
\end{itemize}

These indices are calculated for each journal one by one.
Let $Q_{N,k}$ be the number of citations in year~$N$
of the articles published in the journal in year $N-k$.
By $P_N$ we denote the number of the articles published
by the journal in year~$N$. Note in passing that N
is the number of a year in the

Gregorian calendar,
and so  $N$ is at least six since
$N$ is greater than thousand. In this notation  $\MCQ_N$,
the MCQ of the journal in year~$N$, is calculated as follows:

$$
\MCQ_N =\frac{Q_{N,1} + Q_{N,2} +\ldots + Q_{N,5}}{P_{N-1} + P_{N-2} +\dots+ P_{N-5}}.
$$

Denote the impact factor in year $N$ by $\IF_N$.
By definition
$$
\IF_N =\frac{Q_{N,1} + Q_{N,2}}{P_{N-1}+P_{N-2}}.
$$
Thus, MCQ and IF are defined by the same scheme covering
the different time spans of the relevant databases.\footnote{Cp.~\cite{Weinstock}.}
The first takes the citations of the previous five years; whereas
the second, of the last two years. The RISC and  MNRU impact factors
are calculated by the classical two-year formula for IF suggested by

Eugene Garfield,\footnote{Cp.~\cite{Garfield}.}
 the founder of the Institute for Scientific Information.
It is worth observing that all four indices use different although intersecting
databases.

Let us assume that all articles in some journal
are of the same high quality and has the same number of citations.
Assume further that the number of articles in any volume
is the same every year. In other words,
suppose that $Q_{N,k}$ and $P_N$ are independent of~$N$ and~$k$.
In~this model case, the MCQ and   IF
of the journal must coincide with one another as well as with the remaining
two indices. Fluctuations are inevitable in practical situations, but
the trend to coincidence should prevail for sufficiently full
databases. However, we observe nothing like this
for the real indices. The discrepancies in their
actual values for a particular journal seem improbable for
random fluctuations. For instance, IF is twice as much as MCQ
for a few  outstanding mathematical journals.

By way of illustration let us compare
the current impact factors of the two pairs of prestigious  journals
on algebra and logic:

\smallskip
\hskip2cm\vbox{
\def\tablerule{\noalign{\hrule}}
\halign{\strut\hfil #&\enspace #\hfil&\quad\hfil #&\quad\hfil #\cr
      &                       &   IF & MCQ\cr
    \tablerule
     & J. Algebra             &   0.630 &   0.64\cr
     & J. Pure Appl.Algebra   &   0.666 &   0.59\cr
     & J. Symb. Logic         &   0.609 &   0.31\cr
     & J. Pure Appl. Logic    &   0.613 &   0.30\cr
    \tablerule
     &&&\cr}
}

Using MCQ it is possible to conclude that
the two logical journals are twice as ``feeble'' as their
algebraic counterparts. In fact, the practical coincidence
of the IF and MCQ  of the two algebraic journals
demonstrates  most likely that  the articles of these journals
primarily attract the scientists that publish their papers in the journals
covered by~MR.  At~the same time, more than a half of the citations
of the two logical journals appears in the sources that are not scanned
by~MR. Therefore, the scope of influence of the logical pair on the flux of scientific
information is substantially broader than that
of the other pair. Moreover, the narrow audience is hardly
a merit of any scientific journal.

The differences in databases greatly effect the calculation of the indices of
Russian periodicals.\footnote{Cp.~\cite{Khaitun}.} Let us take a look at the current values of
the above-mentioned indices for a~few authoritative journals of the
Russian Academy of Sciences. The first four of them  publish papers in all areas
of mathematics, and the fifth is interdisciplinary.

\smallskip
\hskip.5cm\vbox{
\def\tablerule{\noalign{\hrule}}
\halign{\strut\hfil #&\enspace #\hfil&\quad\hfil #&\quad\hfil #&\quad\hfil # &\hfil # &\hfil# &\hfil#\cr
      &                      &   IF & MCQ& RISC & MNRU &Founded in\cr
    \tablerule
     & Sb. Math.             &   0.359  &   0.44 &   0.113  &   0.399   &  1866\quad\cr
     & Russ. Math. Surv.     &   0.309  &   0.35 &   0.103  &   0.382   &  1936\quad\cr
     & Sib. Math. J.         &   0.208  &   0.18 &   0.108  &   0.269   &  1960\quad\cr
     & Math. Notes           &   0.251  &   0.18 &   0.030  &   0.244   &  1967\quad\cr
     & Theoret. Math. Phys.  &   0.622  &   0.12 &   0.107  &   0.601   &  1969\quad\cr
    \tablerule
     &&&&&&\cr}
}
\hskip1cm

The obvious conclusion is in order that, taken  {\it per annum},
all indices under consideration  primarily
characterize the respective databases, slightly reflecting
a~minor part of few phenomena of the real functioning of science.

The dynamics of citation indices may be more informative.
For instance, look at the impact factors IF and MCQ of
the

{\it Russian Journal of Mathematical Physics}:

\smallskip
\hskip2cm\vbox{
\def\tablerule{\noalign{\hrule}}
\halign{\strut\hfil #& #\hfil&\quad\hfil #&\quad\hfil #\cr
              & &IF & MCQ\cr
    \tablerule
     & 2003   & 0.291& 0.23\cr
     & 2004   & 0.348& 0.19\cr
     & 2005   & 0.394& 0.26\cr
     & 2006   & 0.493& 0.34\cr
     & 2007   & 1.012& 0.35\cr
    \tablerule
     &&&\cr}
}

Viktor Maslov,

Editor-in-Chief of
this journal, indicates that a few publications on economic applications of the ideas
of mathematical physics might be  a reason for the almost two-times
raise of IF in~2007. Incidentally, MCQ neglects this phenomenon completely.

Traffic congestion never reflects the artistic gifts of  jammed
drivers. By analogy, there are insufficient grounds to correlate
rather arbitrary numerical indices of the dynamics of scientific information
in a particular database with the quality of publications, all mystical hypotheses
of the bureaucracy of science
notwithstanding.

Science is not
the

glass bead game

 despite whatever ciphers.

\bibliographystyle{plain}

\end{document}